\documentclass[prl,reprint,showpacs]{revtex4-1}
\usepackage{color,graphicx}
\usepackage{amssymb,amsmath,amsfonts}
\usepackage{bm}

\renewcommand{\vec}[1]{\bm{#1}}
\newcommand{\Rcal}{\mathcal{R}}

\begin{document}

\title{Shortcut To Adiabaticity for an Anisotropic Gas 
  Containing Quantum Defects}
\author{D.J. Papoular and S. Stringari}
\affiliation{INO-CNR BEC Center and Dipartimento di Fisica, 
  Universit\`a di Trento, 38123 Povo, Italy}
\date{\today}

\begin{abstract}
We present a Shortcut To Adiabaticity (STA) protocol applicable to
3D unitary Fermi gases and 2D
weakly--interacting Bose gases containing defects such as
vortices or solitons. 
Our protocol relies on 
a new class of exact scaling solutions 
in the presence of anisotropic time--dependent harmonic traps. 
It connects stationary states in initial
and final traps having the same frequency ratios. 
The resulting scaling laws exhibit a universal form and also apply to the 
classical Boltzmann gas.
The duration of the STA can be made very short so as to
realize a quantum quench from one stationary state to another.
When applied to an anisotropically trapped  superfluid gas, the STA 
conserves the 
shape of the quantum defects hosted by 
the cloud, 
thereby acting like a perfect microscope,
which sharply constrasts with their strong
distortion occurring  during the free expansion of the cloud.
\end{abstract}


\maketitle

\begin{figure*}
  \centerline{
  \hbox{
    \begin{minipage}{.36\textwidth}
      \includegraphics[width=\textwidth]
      {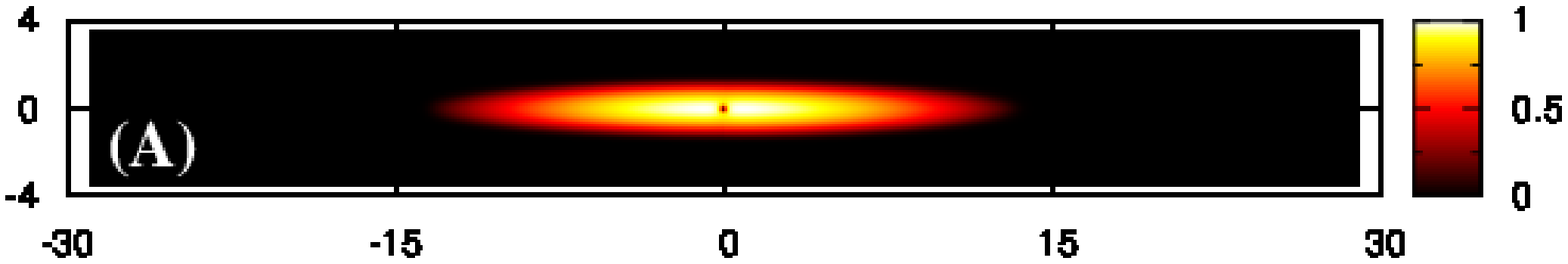}
      \\
      \includegraphics[width=\textwidth]
      {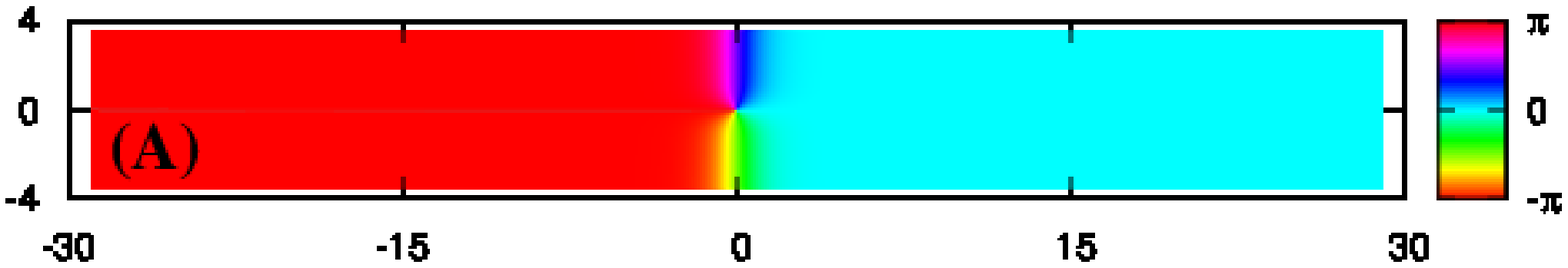}
    \end{minipage}
    \begin{minipage}{.36\textwidth}
      \includegraphics[width=\textwidth]
      {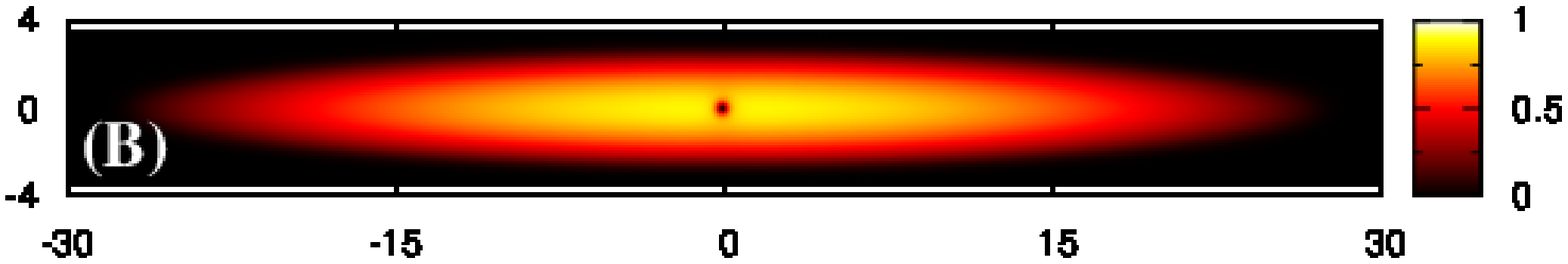}
      \\
      \includegraphics[width=\textwidth]
      {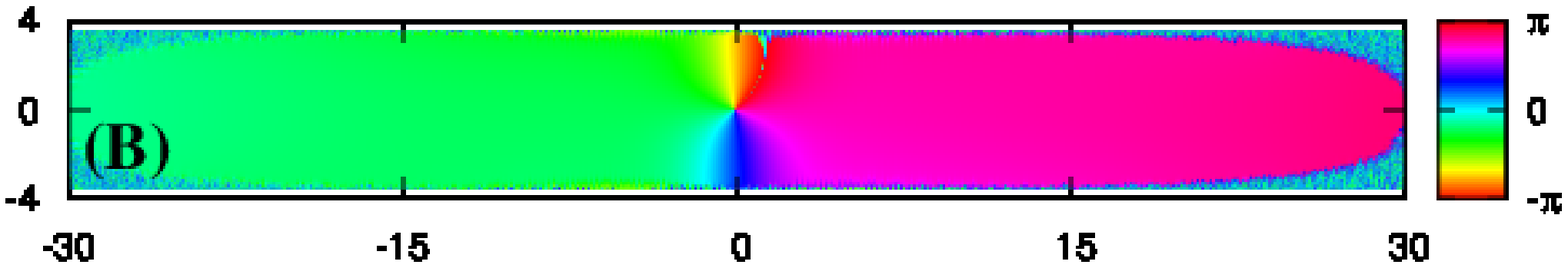}
    \end{minipage}
    \hspace*{.02\textwidth}
    \begin{minipage}{.15\textwidth}
      \centering
      \includegraphics[width=\textwidth]
      {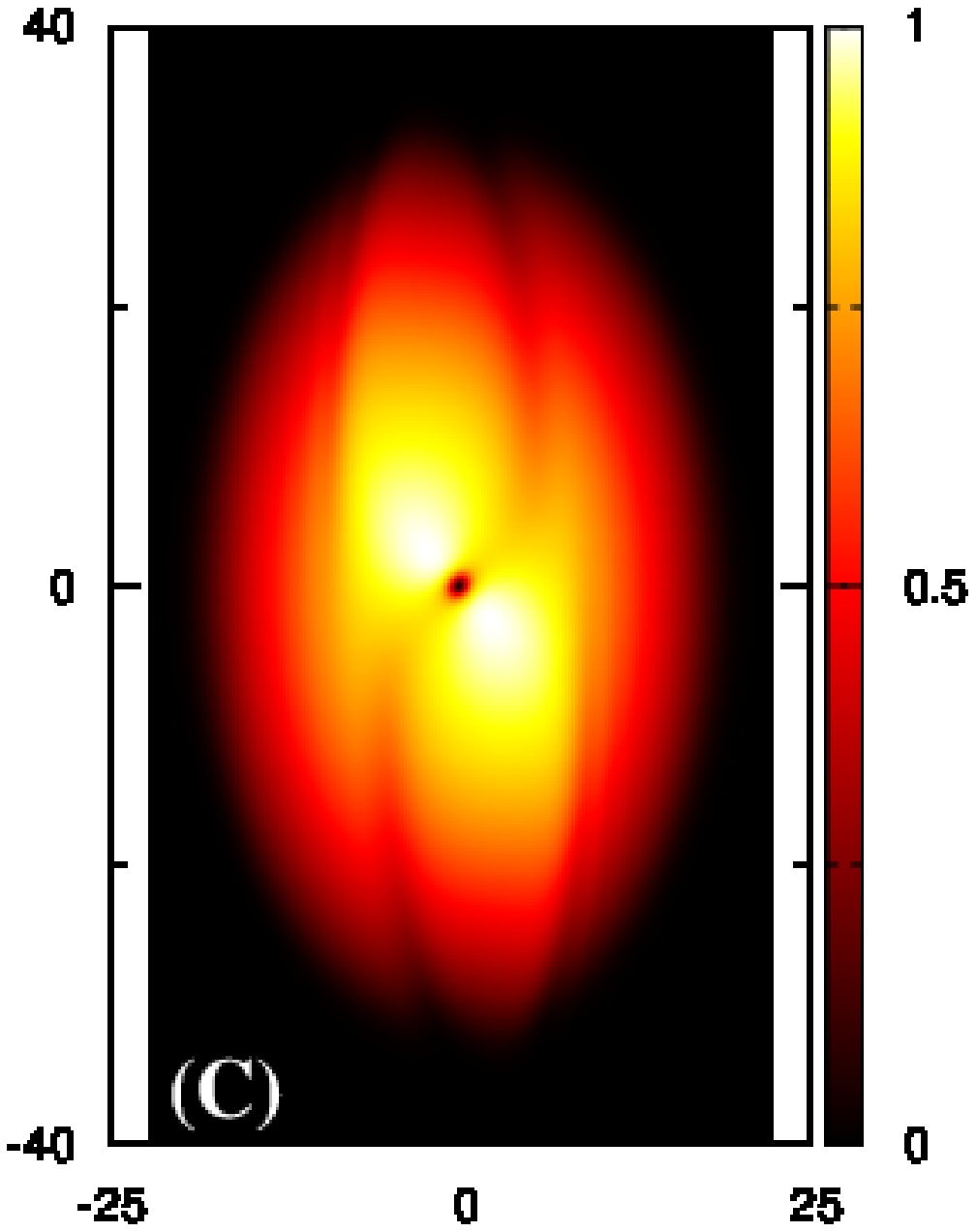}
    \end{minipage}
    \begin{minipage}{.15\textwidth}
      \centering
      \includegraphics[width=\textwidth]
      {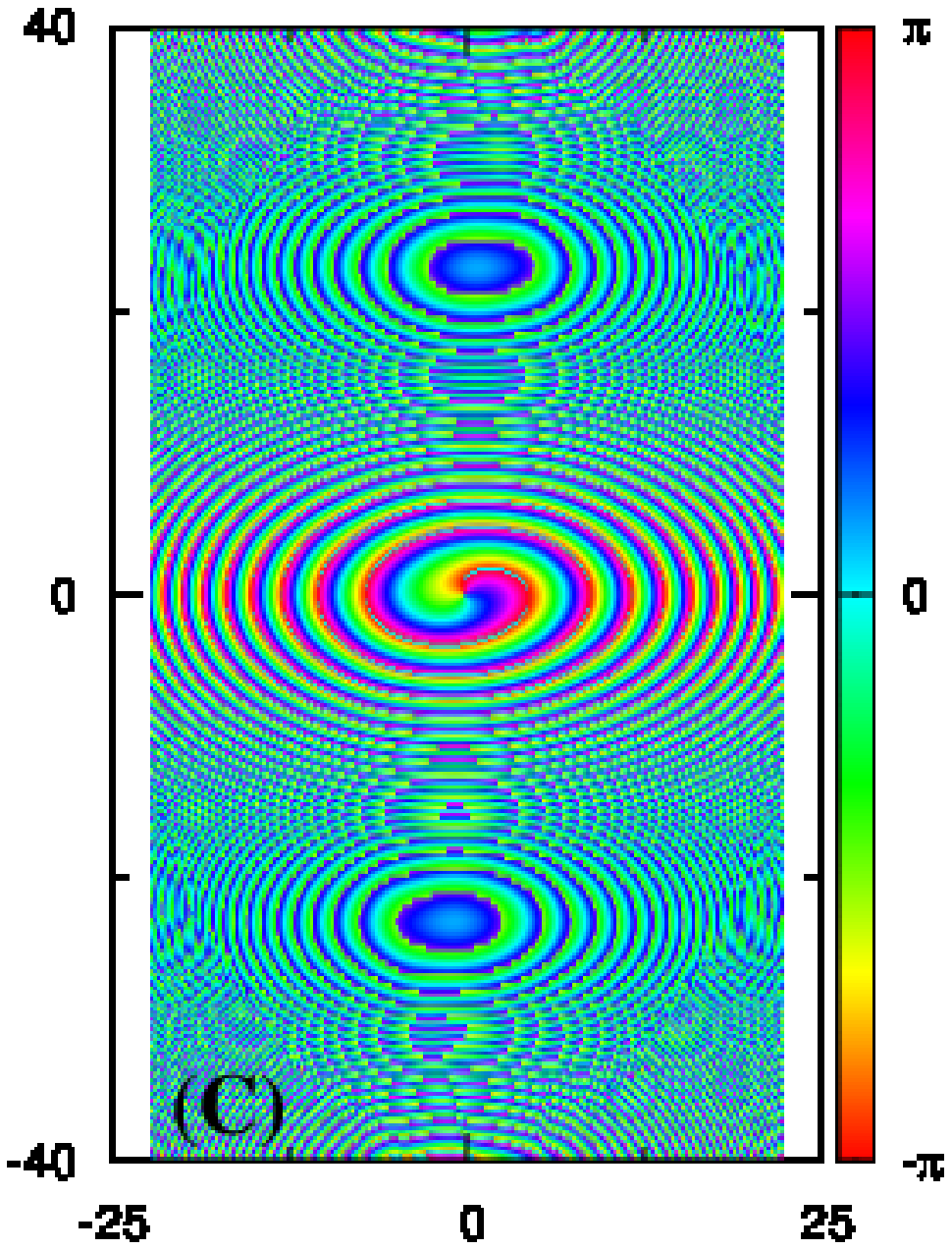}
    \end{minipage}
  }}
  \caption{\label{fig:vortex_STA_freexpand}
  (A) 
  Density (top) and phase (bottom) profiles for an inital anisotropic
  Bose cloud containing a vortex 
  ($\omega_{02}/\omega_{01}=10$ and $mg_\mathrm{2D}N/\hbar^2=3100$,
  corresponding to the chemical potential 
  $\mu_\mathrm{2D}/\hbar\omega_{01}\approx 100$).
  (B)
  Density and phase profiles
  after the STA evolution represented on Fig.~\ref{fig:STA2Danis}A, 
  at the time
  $\omega_{01}t=1.8$. 
  (C) 
  Density and phase profiles after the free expansion of
  the same inital cloud during the same time ($\omega_{01}t=1.8$).
  All six plots result from a numerical solution of 
  Eq.~(\ref{eq:GPE2D}) in imaginary time (left) 
  and real time (center, right).
  Lengths are in units of $(\hbar/m\omega_{01})^{1/2}$ and 
  densities are normalized to the maximum density.}
\end{figure*}

Solitons and quantized vortices are fundamental excitations of
non--linear media and play a key role in superfluid 
dynamics \cite{kevrekidis:Springer2008}.
In anisotropic geometries,
the energetically favored defects are solitonic vortices,
characterized by a non--circular velocity field 
which nevertheless presents non--zero 
circulation \cite{brand:PRA2001,munozmateo:PRL2014}. 
The investigation of their dynamics has been initiated by recent experiments
where they were created
either deterministically by 
phase imprinting \cite{ku:PRL2014}, or spontaneously in a system 
which is quickly
driven through a phase transition \cite{lamporesi:NatPhys2013}.
These defects exhibit intricate dynamics and decay mechanisms. For example,
the snake instability \cite{dutton:Science2001}, 
which affects solitons in 2D and in 3D, can lead
to the creation of a solitonic vortex, a process which is likely to have 
played a role in \cite{donadello:PRL2014}. 
The size of these defects is set by the healing length 
\cite[chap.~5]{blackbook:OUP2003},
which is too small to allow for in--situ observation. 
Up to now they have always been observed after the free expansion of the cloud, 
which increases the core dimensions. 
However, in the anisotropic case,
the free expansion strongly distorts the cloud, and its use in the presence
of a solitonic vortex leads to an involved density profile sporting
a twisted nodal line \cite{tylutki:arXiv2014}.

An alternative to 
free expansion 
is provided
by the recent 
Shortcut To Adiabaticity (STA) schemes 
\cite{chen:PRL2010,torrontegui:AAMO2013},
which reversibly evolve a many--body system from one state to another,
reaching the same target state as an adiabatic transformation 
in a much
shorter time over which decoherence and losses are minimal.
They allow for the manipulation of the momentum spread
of a wavepacket without the time constraints of 
delta kick cooling \cite{ammann:PRL1997,aoki:PRA2006}.
They can be formulated 
as counterdiabatic driving
\cite{delcampo:PRL2013,deffner:PRX2014}.
They are being considered for the preparation of
many--body states 
\cite{delcampo:PRA2011,delcampo:PRL2012,campbell:arXiv2014},
they have motivated an 
exploration of the quantum speed limit
\cite{bason:NatPhys2012} and reflection on the
third law of thermodynamics \cite{levy:PRL2012}.

The construction of an STA  relies on the existence of a scaling
solution to the equation describing the many--body dynamics of the system
in 
a time--dependent trap. 
Such a 
solution exists for the ideal gas in a harmonic trap
and 
has been used to construct an STA solution
\cite{chen:PRL2010} implemented 
experimentally on a 
Bose cloud above $T_c$ \cite{schaff:PRA2010}. 
Allowing for interactions, 
a scaling solution exists for the hydrodynamic equations
holding in the Thomas--Fermi 
regime \cite{kagan:PRA1996,castin:PRL1996,kagan:PRA1997}
and the corresponding STA \cite{muga:JPB2009}
has been realized experimentally
\cite{schaff:EPL2011}. Scaling solutions
also exist for a class of
many--body systems including 
interacting quasi--1D Bose gases 
\cite{gritsev:NJP2010,rohringer:arXiv2013}, 
and for box potentials \cite{delcampo:SciRep2012}.

Up to now, STAs have been demonstrated
either in the ideal--gas  or the Thomas--Fermi regimes. 
Their application to a gas containing 
defects requires going beyond these approximations,
as their existence and dynamics 
result from the interplay between
quantum pressure and interactions.
Harmonically--trapped 2D weakly--interacting Bose gases 
and 3D unitary Fermi gases
are specially promising. 
In the isotropic case, 
they allow for an exact scaling solution to the many--body dynamics
\cite{kagan:PRA1996,castin:Springer2012}
requiring
no time--dependent manipulation of the interactions 
\cite{delcampo:EPL2011},
due to a dynamical 
symmetry 
shared by the 
2D bosonic \cite{pitaevskii:PhysLettA1996,pitaevskii:PRA1997}
and 3D fermionic \cite{castin:Springer2012} 
gases, as well as  by
the classical Boltzmann gas \cite{dgo:PRL2014}.
The absence of
damping of the breathing mode 
due to this symmetry 
has been
confirmed experimentally in the Bose case \cite{chevy:PRL2002}.

In this Letter, we introduce a novel STA protocol applicable
to anisotropic 2D Bose gases 
and 3D unitary Fermi gases
hosting defects, as well as to classical
Boltzmann gases.
Our STA links two stationary states 
in initial
and final traps with the same anisotropy.
It allows for a reversible and
arbitrarily fast compression or expansion of the cloud which conserves
the aspect ratios and acts as a homothety on the defects.
This sharply contrasts with the free expansion of the anisotropic cloud,
which leads to a time--dependent aspect ratio and
an inversion of the cloud anisotropy,
and dramatically affects
the density and phase profiles of vortices \cite{tylutki:arXiv2014}
(see Fig.~\ref{fig:vortex_STA_freexpand}C).
Our 
STA can be used to quench the cloud 
\cite{polkovnikov_RMP2011} from
one anisotropic stationary state to another.
It relies on a new and exact scaling solution for
the dynamics of the cloud  
in a time--dependent anisotropic harmonic trap.
Exact solutions 
for the dynamics of quantum systems 
are rare, 
and we believe ours to be the first analytical solution 
describing the dynamics of a quantum gas
in 
a time--dependent 
anisotropic trap.
It
is applicable 
if the spatial aspect ratio of the cloud remains constant 
in time,
though the 
ratios of the trapping frequencies 
need not be constant.

{\em Scaling solution for 3D anisotropic Fermi gases.} ---
We consider a 3D unitary Fermi gas 
in the harmonic trap
$V(\vec{r},t)=m\sum_{j=1}^D \omega_j^2(t) x_j^2/2$,
where $m$ is the atomic mass and $D=3$ is the number 
of spatial dimensions.
Note that $V(\vec{r},t)$ is both time--dependent and anisotropic.
We describe the system dynamics 
using the Schr\"odinger equation
applied to $N$--particle wavefunctions 
$\Psi(\vec{R})$, 
where
$\vec{R}=(\vec{r}_1,\ldots,\vec{r}_N)$ is the set of all 
particle
coordinates.
Interactions are included through the Bethe--Peierls (BP) boundary
conditions \cite{castin:Springer2012}.
We assume that, for $t\leq 0$, $\Psi$ is a stationary state
(which need not be the ground state 
and may host a defect, e.g., a quantized vortex)
in the constant trap
$V_0(\vec{r})=m\sum_{j=1}^D\omega_{0j}^2x_j^2/2$.
We seek 
a solution  to the Schr\"odinger equation 
of the form \cite{castin:Springer2012}:
\begin{equation} \label{eq:scalingsol_Fermi}
  \Psi(\vec{R},t)
  =\frac{e^{-iE\tau/\hbar}}{b^{ND/2}}
  \exp
  \left[
    \frac{im\dot{b}}{2\hbar b}R^2
  \right]
  \Psi(\vec{R}/b,0)
  \ ,
\end{equation}
where
$b(t)$ is the only scaling parameter, and
$\tau(t)=\int^t dt'/b^2(t')$ is the reduced time. 
The positive function $b(t)$ satisfies
the boundary conditions $b(0)=1$, $\dot{b}(0)=0$, $\ddot{b}(0)=0$,
where the dots denote derivation with respect to the time $t$.
We find that, if the trap frequencies satisfy:
\begin{equation} \label{eq:om1om2_forscaling}
  \omega_j^2(t)=\frac{\omega_{0j}^2}{b^4}-\frac{\ddot{b}}{b}
  \quad\text{for $1\leq j\leq D$,}
\end{equation}
then Eq.~(\ref{eq:scalingsol_Fermi}) is an exact solution
of the Schr\"odinger equation for $t>0$. If
$\omega_{01}=\omega_{02}=\omega_{03}$, our solution reduces to
the isotropic solution of Ref.~\cite{castin:Springer2012}.
In the general case, 
the scaling parameter
$b(t)$ being equal along all spatial directions follows
from the requirement that 
Eq.~(\ref{eq:scalingsol_Fermi}) 
conserve the 
BP conditions.
As we shall show below, the condition on the 
$\omega^2_j(t)$'s
for the anisotropic scaling solution to hold 
(Eq.~\ref{eq:om1om2_forscaling})
is 
the same for
all three considered systems (Fermi, Bose, and Boltzmann gases),
and  follows from their shared dynamical symmetry. 
A remarkable feature of the unitary Fermi gas is that the same 
scaling solution applies to all eigenstates of the Hamiltonian 
satisfying the BP conditions, and thus holds at all 
temperatures.

The scaling transformation $\Psi(\vec{R}/b,0)/b^{ND/2}$ 
in Eq.~(\ref{eq:scalingsol_Fermi}) dictates the two accompanying
phase factors. The first term, proportional to $\tau$,
represents the evolution with the rescaled time 
$\tau$ of the phase of a stationary state. 
The second one, proportional to $R^2$, is a gauge transform
following from the continuity equation.
The joint rescaling of the space and 
time coordinates, along with this gauge transform,
allow us to describe the dynamics of the system exactly
using the differential equations 
of Eq.~(\ref{eq:om1om2_forscaling}). 
In particular,
Eq.~(\ref{eq:scalingsol_Fermi}) shows that
all mean--squared radii
of the gas satisfy 
$\langle x_i^2(t)\rangle/ \langle x_i^2\rangle_0
=b^2(t)$, 
where the average
$\langle x_i^2\rangle_0$ 
relates to the stationary configuration for $t\leq 0$.

\begin{figure*}
  \begin{minipage}{.3\textwidth}
    \includegraphics[angle=-90,width=\textwidth]
    {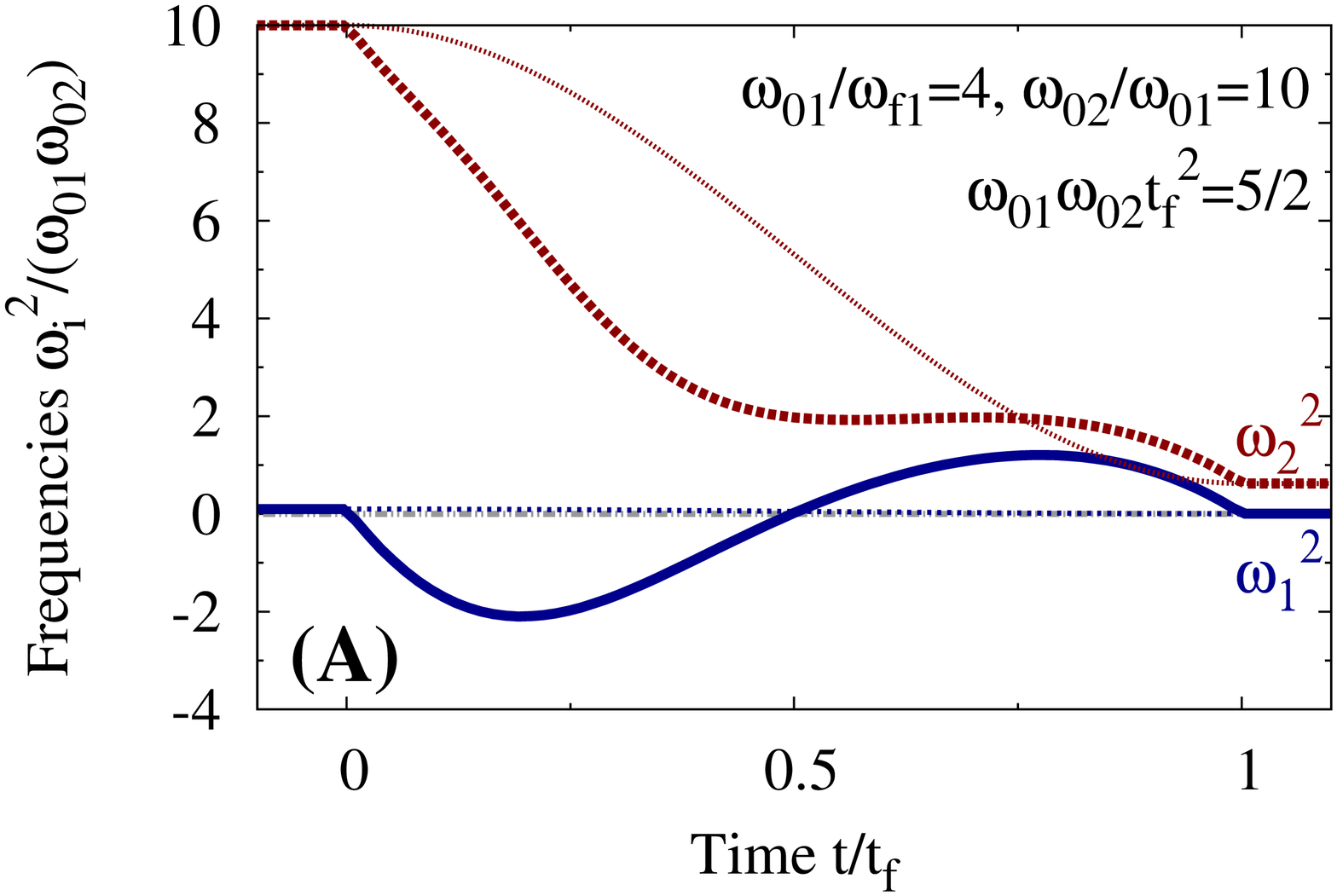}
  \end{minipage}
  \begin{minipage}{.3\textwidth}
    \includegraphics[angle=-90,width=\textwidth]
    {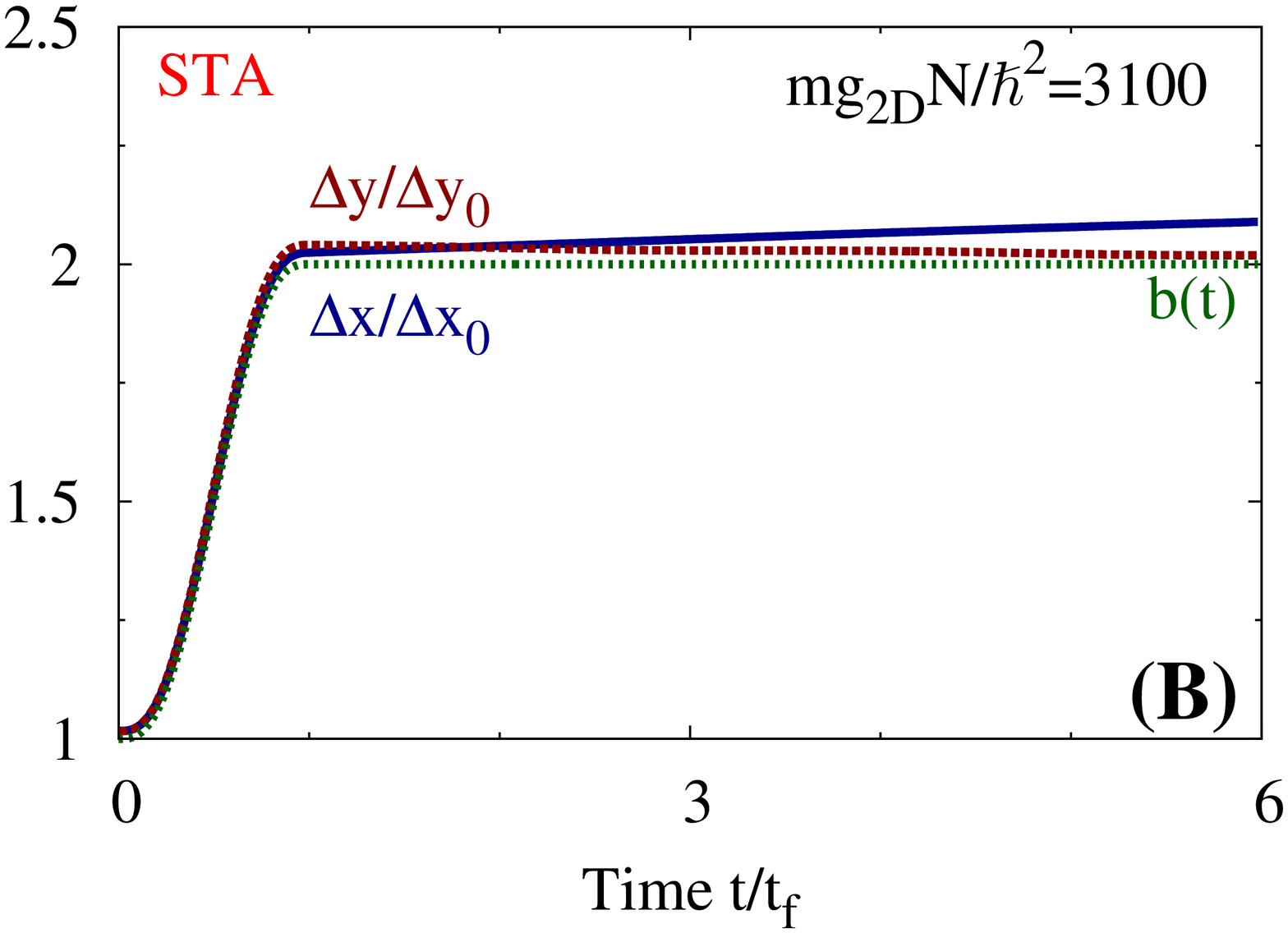}
  \end{minipage}
  \begin{minipage}{.3\textwidth}
    \includegraphics[angle=-90,width=\textwidth]
    {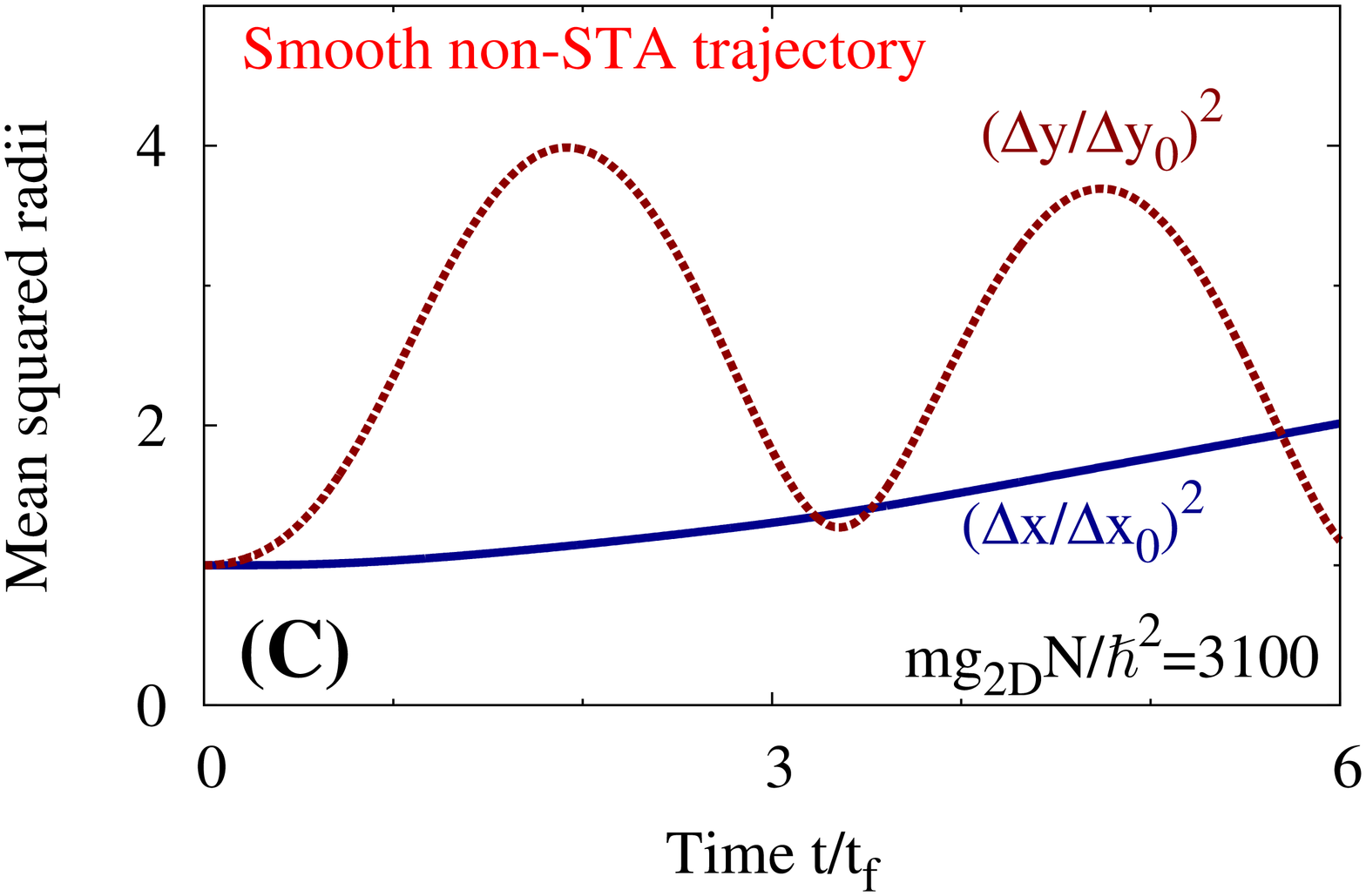}
  \end{minipage}
  \caption{\label{fig:STA2Danis}
     (A)   
     Thick lines:
     Squared trapping frequencies which achieve an anisotropic STA
     linking the stationary states in two 2D traps 
     with $\omega_{02}/\omega_{01}=\omega_{f2}/\omega_{f1}=10$ and 
     $\omega_{01}t_f=0.5$.
     Thin lines: smooth non--STA trajectory.
     (B) 
     Comparison between the expected scaling parameter 
     and the calculated quadratic mean radii 
     $\Delta x=\sqrt{\langle x^2(t)\rangle}$ and
     $\Delta y=\sqrt{\langle y^2(t)\rangle}$
     for the STA trajectory starting from the initial cloud of
     Fig.~\ref{fig:vortex_STA_freexpand}(left)
     for times up to $t/t_f=6$ (i.e. $\omega_{01}t=3$). 
     (C)
     Quadratic mean radii for the non--STA trajectory 
     of panel (A).
     The graphs (B) and (C) result from a numerical solution
     of the GP Eq.~(\ref{eq:GPE2D}) with $mg_\mathrm{2D}N/\hbar^2=3100$.
}
\end{figure*}

{\em 2D Bose gas} ---
We consider a Bose gas which is tightly confined in the 
direction $z$, and trapped in the harmonic potential 
$V(\vec{r},t)=m(\omega_1^2(t) x^2+\omega_2^2(t) y^2)/2$
in the 
directions $x$ and $y$. 
We assume that the oscillator length $l_z$
for the axial confinement
is both larger than the scattering length $a_\mathrm{3D}$ 
characterizing 3D collisions and smaller than the healing length.
The first assumption ensures that the scattering amplitude is 
momentum--independent and that the quantum depletion is negligible,
hence avoiding the quantum anomaly affecting the scale
invariance of 2D systems 
\cite{olshanii:PRL2010,hu:PRL2011,hofmann:PRL2012}, 
while the second 
means that the atomic density does not spill beyond the axial ground state
\cite{merloti:PRA2013}. 
These experimentally realistic conditions 
\cite{hadzibabic:NuovoCimento2011}
allow for a description of 
the $T=0$ dynamics of the system using the 2D Gross--Pitaevskii
equation (GPE) \cite[chap.~17]{blackbook:OUP2003}: 
\begin{equation} \label{eq:GPE2D}
  i\hbar\frac{\partial\Phi}{\partial t}=
  -\frac{\hbar^2}{2m}\Delta\Phi
  +V(\vec{r},t)\Phi
  +g_\mathrm{2D}|\Phi|^2\Phi
  \ ,
\end{equation}
where 
$\Phi(\vec{r},t)$ is the Bose order parameter and
the Laplacian
$\Delta=\partial_x^2+\partial_y^2$.
The coupling constant $g_\mathrm{2D}$, satisfying 
$g_\mathrm{2D}m/\hbar^2=\sqrt{8\pi}a_\mathrm{3D}/l_z$
\cite{petrov:PRA2001}, should be small for
the logarithmic corrections to the equation
of state to be negligible \cite{mora:PRL2009}
and, hence, 
to avoid the quantum anomaly discussed above.
If 
$\omega_{1,2}$ 
fulfill
Eq.~(\ref{eq:om1om2_forscaling}), then
the order parameter 
$\Phi(\vec{r},t)$ in
Eq.~(\ref{eq:GPE2D}) obeys the same scaling 
law Eq.~(\ref{eq:scalingsol_Fermi}) as 
the unitary Fermi gas, with $D=2$ and $N=1$.
The spatial aspect ratio 
$\Rcal=\sqrt{\langle y^2\rangle/\langle x^2\rangle}$ 
of the cloud is constant in time,
even though the frequency ratio $\omega_2(t)/\omega_1(t)$ 
depends on time (Eq.~\ref{eq:om1om2_forscaling}). 
If $\omega_{01}=\omega_{02}$, our solution reduces
to the known isotropic solution \cite{kagan:PRA1996}. 

{\em Boltzmann gas} ---
We finally describe the classical Boltzmann gas through its 
distribution function
$f(\vec{r},\vec{v},t)$ 
governed by
the Boltzmann equation \cite[chap.~2]{cercignani:Springer1988}:
\begin{equation}
  \label{eq:boltzmann}
  \partial_t f
  +\vec{v}\cdot\nabla_{\vec{r}} f
  +\frac{\vec{F}}{m}\cdot\nabla_{\vec{v}} f
  =I_\mathrm{coll}[f]
  \ ,
\end{equation}
where $\vec{F}=-\nabla_{\vec{r}}V$ is the external force,
and $I_\mathrm{coll}[f]$ is the collisional integral.
We seek an $f(\vec{r},\vec{v},t)$ of  the form:
\begin{equation} \label{eq:boltzmann_distrib}
f 
=
  \exp 
  \left[
    -m\beta
    \left(
      \frac{\beta_0^2}{\beta^2}\sum_{i=1}^D\omega_{0i}^2 x_i^2
      +(\vec{v}-{\dot{\beta}\vec{r}}/{2\beta})^2
    \right)
  \right]
\ ,
\end{equation}
where $\beta$ is the inverse temperature
and $\beta_0$ is 
its stationary $t<0$ value.
If the $\omega_i$'s satisfy 
Eq.~(\ref{eq:om1om2_forscaling}) with $b=({\beta/\beta_0})^{1/2}$,
then $f$ is an exact solution of the Boltzmann
Eq.~(\ref{eq:boltzmann}). 
The presence of 
a single scaling parameter reflects the equality of
temperature along all spatial directions.
In the isotropic case, 
Eq.~(\ref{eq:boltzmann_distrib}) reduces to the solution
of Ref.~\cite{dgo:PRL2014}.

{\em Shortcut to adiabaticity} ---
We now use the 
scaling solution described above to 
explicitly construct  an STA 
transforming the initial 
stationary state 
into another stationary state 
in a trap with
the same frequency ratios, i.e.\ 
$\omega_{fi}/\omega_{fj}=\omega_{0i}/\omega_{0j}$. 
The $\omega_{fj}$'s 
describe
the final state, reached 
at the time $t_f$ which is chosen at will.
In analogy with the isotropic scheme of Ref.~\cite{dgo:PRL2014},
we first choose an appropriate
time dependence for $b(t)$, and then  
deduce the 
time--dependent 
frequencies $\{ \omega_j(t) \}$. 
We take $\omega_j(t\geq t_f)=\omega_{fj}$, 
and require
that $\Psi(\vec{r},t)$ be stationary for $t\geq t_f$, 
i.e.\ 
$\dot{b}(t_f)=0$ and $\ddot{b}(t_f)=0$. 
Combined with Eq.~(\ref{eq:om1om2_forscaling}), the latter condition
defines the 
value $b_f=b(t_f)$: 
\begin{equation} \label{eq:cond_w1fw2f}
  \frac{\omega_{fj}}{\omega_{0j}}
  =\frac{1}{b^2(t_f)} 
  \quad \text{for $1\leq j\leq D$.}
\end{equation}
Provided that Eq.~(\ref{eq:cond_w1fw2f}) holds, 
the scaling parameter $b(t)$ must 
satisfy 3 boundary conditions for $t=0$ and 3 others for $t=t_f$
\footnote{The first three conditions 
($b(0)=1$, $\dot{b}(0)=0$, $\ddot{b}(0)=0$)
state the stationarity of $\Psi(\vec{r},t)$ for times $t<0$.
Conditions 4 and 5
($\dot{b}(t_f)=0$, $\ddot{b}(t_f)=0$)
state its stationarity for times $t>t_f$.
Condition 6 follows from Eq.~(\ref{eq:cond_w1fw2f}) and sets the final
value of the scaling parameter ($b(t_f)=({\omega_{01}/\omega_{f1}})^{1/2}$).
}.
Numerous choices are possible for $b(t)$, 
the simplest being
the following fifth--order polynomial:
\begin{equation} \label{eq:bpoly5}
  b(t)=1+(b_f-1)
  \left[
    10\left(\frac{t}{t_f}\right)^3
    -15\left(\frac{t}{t_f}\right)^4
    +6\left(\frac{t}{t_f}\right)^5
  \right]
  \ .
\end{equation}
Once $b(t)$ is known, the frequencies $\omega_j(t)$ 
achieving
the STA are determined using
Eq.~(\ref{eq:om1om2_forscaling}).
Their evolution in time 
does not depend on the interaction strength or the atom number. 
For given values of 
$m$ and $a_{3D}$, the STA trajectory
$\{\omega_j^2(t)\}$ 
depends on
three 
parameters.
First, the scaled duration 
$(\prod\omega_{0j})^{1/D}t_f$ 
determines the attractive or repulsive nature
of the trapping frequencies: 
shorter $t_f$'s require
repulsive potentials 
($\omega_j^2<0$)
for longer fractions of the total time. 
Second, the ratio $\omega_{01}/\omega_{f1}$ sets the final cloud dimensions
(Eq.~\ref{eq:cond_w1fw2f}).
Third, the ratios $\omega_{0i}/\omega_{0j}$ ($i<j$)
set the time--independent spatial
aspect ratios 
$\Rcal_{ij}=[\langle x_j^2 \rangle / \langle x_i^2 \rangle]^{1/2}$. 
Our exact solution holds for all coupling strengths from the ideal gas
($\Rcal_{ij}=(\omega_{0i}/\omega_{0j})^{1/2}$) to the 
Thomas--Fermi regime
($\Rcal_{ij}=\omega_{0i}/\omega_{0j}$).
For $t\geq t_f$, 
$\Psi(\vec{r},t)$ is stationary and its time evolution simply
leads to a phase linear in $(t-t_f)$.

We now demonstrate this new STA protocol 
on the 
expansion of an anisotropic 2D Bose cloud 
containing a single vortex 
(Figs.~\ref{fig:vortex_STA_freexpand} and \ref{fig:STA2Danis}).
The single--vortex state is a stationary state of Eq.~(\ref{eq:GPE2D}). 
It is the lowest--energy state satisfying
the symmetry conditions $\Psi(-x,-y,t)=-\Psi(x,y,t)$ and
$\Psi(-x,y,t)=\Psi^*(x,y,t)$. 
Imposing these conditions on the initial wavefunction
allows for its calculation using
imaginary--time evolution despite the absence of any gauge or rotation
term in Eq.~(\ref{eq:GPE2D}).
We start from the single--vortex stationary state with
$\omega_{02}=10\,\omega_{01}$ and $mgN/\hbar^2=3100$,
whose density and phase profiles 
are represented on Fig.~\ref{fig:vortex_STA_freexpand} (left, top and bottom).
Over the short time
$\omega_{01}t_f=1/2$, the STA trajectory reaches a new stationary state
where both trapping frequencies are four times as small as
their initial values. 
Figure~\ref{fig:STA2Danis}A shows $\omega_{1,2}^2(t)$
for 
$0\leq t \leq t_f$.
The exact prediction for the scaling parameter $b(t)$
(Eq.~(\ref{eq:bpoly5}))
is compared on Fig.~\ref{fig:STA2Danis}B
to the calculated values for the ratio
$\Delta x/\Delta x_0=\Delta y/\Delta y_0$ characterizing the
mean radii of the cloud, obtained through a numerical solution
of the GPE 
Eq.~(\ref{eq:GPE2D}) for times up to
$t=6t_f=3/\omega_{01}$.
The density and phase profiles of the cloud at $t=3.6t_f$, shown on
Fig.~\ref{fig:vortex_STA_freexpand} (center),
are those of a stationary anisotropic vortex in the expanded trap.
The small residual oscillations seen on Fig.~\ref{fig:STA2Danis}B
for $t\gtrsim t_f$, which start before $t_f$ and
survive for longer times, are an artefact of the numerical simulation,
and their amplitude decreases with increasing resolutions of 
the spatial grid.

The above results illustrate three important properties of our STA protocol.
First, the 
$\omega_j^2$'s
determined by Eq.~(\ref{eq:om1om2_forscaling})
can be transiently negative, corresponding to an expulsive potential,
as already noted in Ref.~\cite{dgo:PRL2014}.
Second, despite the initial and final trap anisotropies being the same, 
the ratio $\omega_2^2(t)/\omega_1^2(t)$ is not constant in time.
Indeed, Fig.~\ref{fig:STA2Danis}A shows that 
$\omega_2^2(t)>0$ 
at all times whereas
$\omega_1^2(t)$ is negative for intermediate times,
which is a dramatic consequence of anisotropy. 
Third, the 
$\omega_{j}^2$'s 
are continuous
at $t=0$ and $t=t_f$, but they need not go smoothly to the initial and
final values, as long as 
$b(t)$ satisfies the correct 
boundary conditions.
By contrast, even a very smooth 
non--STA trajectory linking the initial and final states within
the same time $t_f$
results in
large--amplitude oscillations of the cloud radii for $t>t_f$
(Fig.~\ref{fig:STA2Danis}C). 
The discontinuity in the derivatives of $\omega^2_{1,2}$ at
$t=0$ and $t_f$  
follows from our choice for $b(t)$ (Eq.~\ref{eq:bpoly5}), and can be
avoided by choosing a higher--order polynomial. 

\begin{figure}
    \includegraphics[angle=-90,width=.7\linewidth]
    {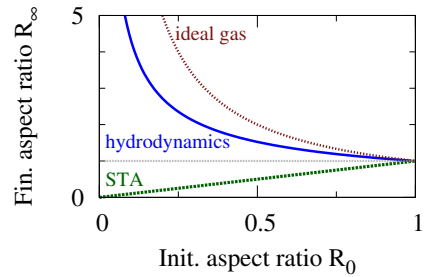}
  \caption{\label{fig:hydro_quadrature}
  Final aspect ratio $\Rcal_\infty$ of the anisotropic cloud 
  as a function of the initial value $\Rcal_0$,
  for the hydrodynamic (solid blue) 
  and ideal--gas ($\Rcal_\infty=1/\Rcal_0$, dotted brown)
  free expansions,
  and for an STA evolution ($\Rcal_\infty=\Rcal_0$, dashed green).
}
\end{figure}

The stationary behavior of the cloud following
an STA (Fig.~\ref{fig:STA2Danis}B) sharply contrasts with its
behavior during free expansion, which
leads to a time--dependent aspect ratio and an 
inversion of the
cloud anisotropy
for all interaction strengths
(Fig.~\ref{fig:hydro_quadrature}).
Unlike the STA,
the free expansion from an anisotropic trap is 
not  governed by a single scaling parameter 
and cannot be formulated in a universal way. In particular,
the scaling parameters $b_j$, characterizing the 
dilation along the spatial directions $1\leq j\leq D$, 
obey different equations in the ideal gas and the hydrodynamic regimes. 
For the 2D ideal Bose gas at $T=0$,
the scaling law reads $\ddot b_j= \omega^2_{0j}/b_j^3$.
Instead, in the hydrodynamic regime, it reads
$\ddot b_j= \omega^2_{0j}/(b_jb_xb_y)$ for the 2D Bose gas 
and $\ddot b_j= \omega^2_{0j}/[b_j(b_xb_yb_z)^{2/3}]$ for the 3D unitary 
Fermi gas \cite{menotti:PRL2002}. Only in the case of isotropic 
trapping do the above laws take the universal 
form $\ddot b= \omega^2_{0}/b^3$ which, by the way, coincides 
with Eq.~(\ref{eq:om1om2_forscaling}) after setting 
$\omega_{0j}=\omega_0$ and $\omega_j(t)=0$ for $t>0$. 
The anisotropy of the hydrodynamic scaling laws 
leads to
a peculiar change in the shape of  the 
expanding density profiles,
which was experimentally observed in \cite{ohara:Science2002}
with a unitary Fermi gas.

Comparing the density and phase profiles obtained 
from the initial cloud containing a vortex and undergoing an STA 
(Fig.~\ref{fig:vortex_STA_freexpand}B) or free expansion
(Fig.~\ref{fig:vortex_STA_freexpand}C) confirms that 
the STA acts as a 
homothety
on the vortex, whereas
free expansion yields the twisted nodal line characterizing
soliton vortices \cite{tylutki:arXiv2014}.
The free expansion of a gas is practically
irreversible. On the contrary,
the time reversal of an STA leads to another STA 
which performs the inverse transformation.
More precisely, if $\{\omega_j(t)\}$ 
is an STA satisfying Eq.~(\ref{eq:om1om2_forscaling})
with the scaling parameter $b(t)$, 
then $\{\omega_j(t_f-t)\}$ is an STA 
for the scaling  function
$\bar{b}(t)=b(t_f-t)/b_f$. Thus, an 
STA compressing the cloud 
is obtained by time--reversing the trajectory 
of Fig.~\ref{fig:STA2Danis}A.

The duration $t_f$ of the STA can be 
arbitrarily small.
Therefore, the STA does not exhibit any quantum 
speed limit \cite{brouzos:arXiv2014}.
If 
$\omega_{f1}/\omega_{01}$ is 
sufficiently small,
$b_f$ can be made large (Eq.~(\ref{eq:cond_w1fw2f})). 
Hence,
the cloud can be expanded to
an arbitrarily large radius within an arbitrarily small time,
thus realizing 
a quantum quench from one stationary state to another,
with the caveat that 
the expulsive part of the trajectory
($\omega_j^2<0$) becomes more pronounced.

In conclusion, our exact scaling solution for Fermi, Bose, and Boltzmann
gases is an important step 
towards the analytical description of the dynamics of 
anisotropic systems. 
The experimental demonstration of the dynamical $\mathrm{SO}(2,1)$ symmetry  
has so far remained elusive in 3D geometries, where 
isotropic traps are 
challenging to achieve, and
the application of our STA to an anisotropic 3D
Fermi gas will provide its 
experimental signature.
Quasi--2D Bose gases are readily created and manipulated
experimentally \cite{hadzibabic:NuovoCimento2011},
and the required anisotropic
(trapping or expulsive) 
potentials  
can be obtained by combining red-- and
blue--detuned optical potentials \cite{he:OptExpress2012}. 
Our protocol
will allow for a detailed experimental investigation of the dynamics 
and decay mechanisms of defects in anisotropic clouds 
\cite{donadello:PRL2014,ku:PRL2014}.
For instance, the 
cloud can be cooled and condensed in a 
strongly--confining trap where
evaporative cooling is efficient, and then quickly expanded using the STA,
without any distortion, into a
weakly--confining trap where defects are easier to observe.

\begin{acknowledgments}
We are grateful to L. Pitaevskii, Y.~Castin, G.~Lamporesi, and H.~Perrin
for fruitful discussions. 
This work has been supported by ERC through the QGBE grant.
\end{acknowledgments}

%\bibliography{STA2Dbib}

\begin{thebibliography}{49}%
\makeatletter
\providecommand \@ifxundefined [1]{%
 \@ifx{#1\undefined}
}%
\providecommand \@ifnum [1]{%
 \ifnum #1\expandafter \@firstoftwo
 \else \expandafter \@secondoftwo
 \fi
}%
\providecommand \@ifx [1]{%
 \ifx #1\expandafter \@firstoftwo
 \else \expandafter \@secondoftwo
 \fi
}%
\providecommand \natexlab [1]{#1}%
\providecommand \enquote  [1]{``#1''}%
\providecommand \bibnamefont  [1]{#1}%
\providecommand \bibfnamefont [1]{#1}%
\providecommand \citenamefont [1]{#1}%
\providecommand \href@noop [0]{\@secondoftwo}%
\providecommand \href [0]{\begingroup \@sanitize@url \@href}%
\providecommand \@href[1]{\@@startlink{#1}\@@href}%
\providecommand \@@href[1]{\endgroup#1\@@endlink}%
\providecommand \@sanitize@url [0]{\catcode `\\12\catcode `\$12\catcode
  `\&12\catcode `\#12\catcode `\^12\catcode `\_12\catcode `\%12\relax}%
\providecommand \@@startlink[1]{}%
\providecommand \@@endlink[0]{}%
\providecommand \url  [0]{\begingroup\@sanitize@url \@url }%
\providecommand \@url [1]{\endgroup\@href {#1}{\urlprefix }}%
\providecommand \urlprefix  [0]{URL }%
\providecommand \Eprint [0]{\href }%
\providecommand \doibase [0]{http://dx.doi.org/}%
\providecommand \selectlanguage [0]{\@gobble}%
\providecommand \bibinfo  [0]{\@secondoftwo}%
\providecommand \bibfield  [0]{\@secondoftwo}%
\providecommand \translation [1]{[#1]}%
\providecommand \BibitemOpen [0]{}%
\providecommand \bibitemStop [0]{}%
\providecommand \bibitemNoStop [0]{.\EOS\space}%
\providecommand \EOS [0]{\spacefactor3000\relax}%
\providecommand \BibitemShut  [1]{\csname bibitem#1\endcsname}%
\let\auto@bib@innerbib\@empty
%</preamble>
\bibitem [{\citenamefont {Kevrekidis}\ \emph {et~al.}(2008)\citenamefont
  {Kevrekidis}, \citenamefont {Frantzeskakis},\ and\ \citenamefont
  {Carretero{-}Gonz{\'a}les}}]{kevrekidis:Springer2008}%
  \BibitemOpen
  \bibinfo {editor} {\bibfnamefont {P.~G.}\ \bibnamefont {Kevrekidis}},
  \bibinfo {editor} {\bibfnamefont {D.~J.}\ \bibnamefont {Frantzeskakis}}, \
  and\ \bibinfo {editor} {\bibfnamefont {R.}~\bibnamefont
  {Carretero{-}Gonz{\'a}les}},\ eds.,\ \href@noop {} {\emph {\bibinfo {title}
  {Emergent non-linear phenomena in {B}ose-{E}instein condensates}}}\ (\bibinfo
   {publisher} {Springer},\ \bibinfo {year} {2008})\BibitemShut {NoStop}%
\bibitem [{\citenamefont {Brand}\ and\ \citenamefont
  {Reinhardt}(2002)}]{brand:PRA2001}%
  \BibitemOpen
  \bibfield  {author} {\bibinfo {author} {\bibfnamefont {J.}~\bibnamefont
  {Brand}}\ and\ \bibinfo {author} {\bibfnamefont {W.~P.}\ \bibnamefont
  {Reinhardt}},\ }\href@noop {} {\bibfield  {journal} {\bibinfo  {journal}
  {Phys. Rev. A}\ }\textbf {\bibinfo {volume} {65}},\ \bibinfo {pages} {043612}
  (\bibinfo {year} {2002})}\BibitemShut {NoStop}%
\bibitem [{\citenamefont {{Munoz Mateo}}\ and\ \citenamefont
  {Brand}(2014)}]{munozmateo:PRL2014}%
  \BibitemOpen
  \bibfield  {author} {\bibinfo {author} {\bibfnamefont {A.}~\bibnamefont
  {{Munoz Mateo}}}\ and\ \bibinfo {author} {\bibfnamefont {J.}~\bibnamefont
  {Brand}},\ }\href@noop {} {\bibfield  {journal} {\bibinfo  {journal} {Phys.
  Rev. Lett.}\ }\textbf {\bibinfo {volume} {113}},\ \bibinfo {pages} {255302}
  (\bibinfo {year} {2014})}\BibitemShut {NoStop}%
\bibitem [{\citenamefont {Ku}\ \emph {et~al.}(2014)\citenamefont {Ku},
  \citenamefont {Ji}, \citenamefont {Mukherjee}, \citenamefont
  {Guardado{-}Sanchez}, \citenamefont {Cheuk}, \citenamefont {Yefsah},\ and\
  \citenamefont {Zwierlein}}]{ku:PRL2014}%
  \BibitemOpen
  \bibfield  {author} {\bibinfo {author} {\bibfnamefont {M.~J.~H.}\
  \bibnamefont {Ku}}, \bibinfo {author} {\bibfnamefont {W.}~\bibnamefont {Ji}},
  \bibinfo {author} {\bibfnamefont {B.}~\bibnamefont {Mukherjee}}, \bibinfo
  {author} {\bibfnamefont {E.}~\bibnamefont {Guardado{-}Sanchez}}, \bibinfo
  {author} {\bibfnamefont {L.~W.}\ \bibnamefont {Cheuk}}, \bibinfo {author}
  {\bibfnamefont {T.}~\bibnamefont {Yefsah}}, \ and\ \bibinfo {author}
  {\bibfnamefont {M.~W.}\ \bibnamefont {Zwierlein}},\ }\href@noop {} {\bibfield
   {journal} {\bibinfo  {journal} {Phys. Rev. Lett.}\ }\textbf {\bibinfo
  {volume} {113}},\ \bibinfo {pages} {065301} (\bibinfo {year}
  {2014})}\BibitemShut {NoStop}%
\bibitem [{\citenamefont {Lamporesi}\ \emph {et~al.}(2013)\citenamefont
  {Lamporesi}, \citenamefont {Donadello}, \citenamefont {Serafini},
  \citenamefont {Dalfovo},\ and\ \citenamefont
  {Ferrari}}]{lamporesi:NatPhys2013}%
  \BibitemOpen
  \bibfield  {author} {\bibinfo {author} {\bibfnamefont {G.}~\bibnamefont
  {Lamporesi}}, \bibinfo {author} {\bibfnamefont {S.}~\bibnamefont
  {Donadello}}, \bibinfo {author} {\bibfnamefont {S.}~\bibnamefont {Serafini}},
  \bibinfo {author} {\bibfnamefont {F.}~\bibnamefont {Dalfovo}}, \ and\
  \bibinfo {author} {\bibfnamefont {G.}~\bibnamefont {Ferrari}},\ }\href@noop
  {} {\bibfield  {journal} {\bibinfo  {journal} {Nat. Phys.}\ }\textbf
  {\bibinfo {volume} {9}},\ \bibinfo {pages} {656} (\bibinfo {year}
  {2013})}\BibitemShut {NoStop}%
\bibitem [{\citenamefont {Dutton}\ \emph {et~al.}(2001)\citenamefont {Dutton},
  \citenamefont {Budde}, \citenamefont {Slowe},\ and\ \citenamefont
  {Hau}}]{dutton:Science2001}%
  \BibitemOpen
  \bibfield  {author} {\bibinfo {author} {\bibfnamefont {Z.}~\bibnamefont
  {Dutton}}, \bibinfo {author} {\bibfnamefont {M.}~\bibnamefont {Budde}},
  \bibinfo {author} {\bibfnamefont {C.}~\bibnamefont {Slowe}}, \ and\ \bibinfo
  {author} {\bibfnamefont {L.~V.}\ \bibnamefont {Hau}},\ }\href@noop {}
  {\bibfield  {journal} {\bibinfo  {journal} {Science}\ }\textbf {\bibinfo
  {volume} {293}},\ \bibinfo {pages} {663} (\bibinfo {year}
  {2001})}\BibitemShut {NoStop}%
\bibitem [{\citenamefont {Donadello}\ \emph {et~al.}(2014)\citenamefont
  {Donadello}, \citenamefont {Serafini}, \citenamefont {Tylutki}, \citenamefont
  {Pitaevskii}, \citenamefont {Dalfovo}, \citenamefont {Lamporesi},\ and\
  \citenamefont {Ferrari}}]{donadello:PRL2014}%
  \BibitemOpen
  \bibfield  {author} {\bibinfo {author} {\bibfnamefont {S.}~\bibnamefont
  {Donadello}}, \bibinfo {author} {\bibfnamefont {S.}~\bibnamefont {Serafini}},
  \bibinfo {author} {\bibfnamefont {M.}~\bibnamefont {Tylutki}}, \bibinfo
  {author} {\bibfnamefont {L.~P.}\ \bibnamefont {Pitaevskii}}, \bibinfo
  {author} {\bibfnamefont {F.}~\bibnamefont {Dalfovo}}, \bibinfo {author}
  {\bibfnamefont {G.}~\bibnamefont {Lamporesi}}, \ and\ \bibinfo {author}
  {\bibfnamefont {G.}~\bibnamefont {Ferrari}},\ }\href@noop {} {\bibfield
  {journal} {\bibinfo  {journal} {Phys. Rev. Lett.}\ }\textbf {\bibinfo
  {volume} {113}},\ \bibinfo {pages} {065302} (\bibinfo {year}
  {2014})}\BibitemShut {NoStop}%
\bibitem [{\citenamefont {Pitaevskii}\ and\ \citenamefont
  {Stringari}(2003)}]{blackbook:OUP2003}%
  \BibitemOpen
  \bibfield  {author} {\bibinfo {author} {\bibfnamefont {L.}~\bibnamefont
  {Pitaevskii}}\ and\ \bibinfo {author} {\bibfnamefont {S.}~\bibnamefont
  {Stringari}},\ }\href@noop {} {\emph {\bibinfo {title} {{B}ose-{E}instein
  Condensation}}}\ (\bibinfo  {publisher} {Oxford University Press},\ \bibinfo
  {year} {2003})\BibitemShut {NoStop}%
\bibitem [{\citenamefont {Tylutki}\ \emph {et~al.}(2014)\citenamefont
  {Tylutki}, \citenamefont {Donadello}, \citenamefont {Serafini}, \citenamefont
  {Pitaevskii}, \citenamefont {Dalfovo}, \citenamefont {Lamporesi},\ and\
  \citenamefont {Ferrari}}]{tylutki:arXiv2014}%
  \BibitemOpen
  \bibfield  {author} {\bibinfo {author} {\bibfnamefont {M.}~\bibnamefont
  {Tylutki}}, \bibinfo {author} {\bibfnamefont {S.}~\bibnamefont {Donadello}},
  \bibinfo {author} {\bibfnamefont {S.}~\bibnamefont {Serafini}}, \bibinfo
  {author} {\bibfnamefont {L.~P.}\ \bibnamefont {Pitaevskii}}, \bibinfo
  {author} {\bibfnamefont {F.}~\bibnamefont {Dalfovo}}, \bibinfo {author}
  {\bibfnamefont {G.}~\bibnamefont {Lamporesi}}, \ and\ \bibinfo {author}
  {\bibfnamefont {G.}~\bibnamefont {Ferrari}},\ }\href@noop {} {\bibfield
  {journal} {\bibinfo  {journal} {arXiv:1410.5475}\ } (\bibinfo {year}
  {2014})}\BibitemShut {NoStop}%
\bibitem [{\citenamefont {Chen}\ \emph {et~al.}(2010)\citenamefont {Chen},
  \citenamefont {Ruschhaupt}, \citenamefont {Schmidt}, \citenamefont
  {{delCampo}}, \citenamefont {{G}u{\'e}ry{-}{O}delin},\ and\ \citenamefont
  {Muga}}]{chen:PRL2010}%
  \BibitemOpen
  \bibfield  {author} {\bibinfo {author} {\bibfnamefont {X.}~\bibnamefont
  {Chen}}, \bibinfo {author} {\bibfnamefont {A.}~\bibnamefont {Ruschhaupt}},
  \bibinfo {author} {\bibfnamefont {S.}~\bibnamefont {Schmidt}}, \bibinfo
  {author} {\bibfnamefont {A.}~\bibnamefont {{delCampo}}}, \bibinfo {author}
  {\bibfnamefont {D.}~\bibnamefont {{G}u{\'e}ry{-}{O}delin}}, \ and\ \bibinfo
  {author} {\bibfnamefont {J.~G.}\ \bibnamefont {Muga}},\ }\href@noop {}
  {\bibfield  {journal} {\bibinfo  {journal} {Phys. Rev. Lett.}\ }\textbf
  {\bibinfo {volume} {104}},\ \bibinfo {pages} {063002} (\bibinfo {year}
  {2010})}\BibitemShut {NoStop}%
\bibitem [{\citenamefont {Torrontegui}\ \emph {et~al.}(2013)\citenamefont
  {Torrontegui}, \citenamefont {Ibáñez}, \citenamefont {Martínez-Garaot},
  \citenamefont {Modugno}, \citenamefont {del Campo}, \citenamefont
  {Gu{\'e}ry-Odelin}, \citenamefont {Ruschhaupt}, \citenamefont {Chen},\ and\
  \citenamefont {Muga}}]{torrontegui:AAMO2013}%
  \BibitemOpen
  \bibfield  {author} {\bibinfo {author} {\bibfnamefont {E.}~\bibnamefont
  {Torrontegui}}, \bibinfo {author} {\bibfnamefont {S.}~\bibnamefont
  {Ibáñez}}, \bibinfo {author} {\bibfnamefont {S.}~\bibnamefont
  {Martínez-Garaot}}, \bibinfo {author} {\bibfnamefont {M.}~\bibnamefont
  {Modugno}}, \bibinfo {author} {\bibfnamefont {A.}~\bibnamefont {del Campo}},
  \bibinfo {author} {\bibfnamefont {D.}~\bibnamefont {Gu{\'e}ry-Odelin}},
  \bibinfo {author} {\bibfnamefont {A.}~\bibnamefont {Ruschhaupt}}, \bibinfo
  {author} {\bibfnamefont {X.}~\bibnamefont {Chen}}, \ and\ \bibinfo {author}
  {\bibfnamefont {J.~G.}\ \bibnamefont {Muga}},\ }in\ \href@noop {} {\emph
  {\bibinfo {booktitle} {Advances in Atomic, Molecular, and Optical
  Physics}}},\ Vol.~\bibinfo {volume} {62}\ (\bibinfo  {publisher} {Academic
  Press},\ \bibinfo {year} {2013})\ pp.\ \bibinfo {pages} {117 --
  169}\BibitemShut {NoStop}%
\bibitem [{\citenamefont {Ammann}\ and\ \citenamefont
  {Christensen}(1997)}]{ammann:PRL1997}%
  \BibitemOpen
  \bibfield  {author} {\bibinfo {author} {\bibfnamefont {H.}~\bibnamefont
  {Ammann}}\ and\ \bibinfo {author} {\bibfnamefont {N.}~\bibnamefont
  {Christensen}},\ }\href@noop {} {\bibfield  {journal} {\bibinfo  {journal}
  {Phys. Rev. Lett.}\ }\textbf {\bibinfo {volume} {78}},\ \bibinfo {pages}
  {2088} (\bibinfo {year} {1997})}\BibitemShut {NoStop}%
\bibitem [{\citenamefont {Aoki}\ \emph {et~al.}(2006)\citenamefont {Aoki},
  \citenamefont {Kato}, \citenamefont {Tanami},\ and\ \citenamefont
  {Nakamatsu}}]{aoki:PRA2006}%
  \BibitemOpen
  \bibfield  {author} {\bibinfo {author} {\bibfnamefont {T.}~\bibnamefont
  {Aoki}}, \bibinfo {author} {\bibfnamefont {T.}~\bibnamefont {Kato}}, \bibinfo
  {author} {\bibfnamefont {Y.}~\bibnamefont {Tanami}}, \ and\ \bibinfo {author}
  {\bibfnamefont {H.}~\bibnamefont {Nakamatsu}},\ }\href@noop {} {\bibfield
  {journal} {\bibinfo  {journal} {Phys. Rev. A}\ }\textbf {\bibinfo {volume}
  {73}},\ \bibinfo {pages} {063603} (\bibinfo {year} {2006})}\BibitemShut
  {NoStop}%
\bibitem [{\citenamefont {del Campo}(2013)}]{delcampo:PRL2013}%
  \BibitemOpen
  \bibfield  {author} {\bibinfo {author} {\bibfnamefont {A.}~\bibnamefont {del
  Campo}},\ }\href@noop {} {\bibfield  {journal} {\bibinfo  {journal} {Phys.
  Rev. Lett.}\ }\textbf {\bibinfo {volume} {111}},\ \bibinfo {pages} {100502}
  (\bibinfo {year} {2013})}\BibitemShut {NoStop}%
\bibitem [{\citenamefont {Deffner}\ \emph {et~al.}(2014)\citenamefont
  {Deffner}, \citenamefont {Jarzynski},\ and\ \citenamefont {del
  Campo}}]{deffner:PRX2014}%
  \BibitemOpen
  \bibfield  {author} {\bibinfo {author} {\bibfnamefont {S.}~\bibnamefont
  {Deffner}}, \bibinfo {author} {\bibfnamefont {C.}~\bibnamefont {Jarzynski}},
  \ and\ \bibinfo {author} {\bibfnamefont {A.}~\bibnamefont {del Campo}},\
  }\href@noop {} {\bibfield  {journal} {\bibinfo  {journal} {Phys. Rev. X}\
  }\textbf {\bibinfo {volume} {4}},\ \bibinfo {pages} {021013} (\bibinfo {year}
  {2014})}\BibitemShut {NoStop}%
\bibitem [{\citenamefont {del Campo}(2011{\natexlab{a}})}]{delcampo:PRA2011}%
  \BibitemOpen
  \bibfield  {author} {\bibinfo {author} {\bibfnamefont {A.}~\bibnamefont {del
  Campo}},\ }\href@noop {} {\bibfield  {journal} {\bibinfo  {journal} {Phys.
  Rev. A}\ }\textbf {\bibinfo {volume} {84}},\ \bibinfo {pages} {031606(R)}
  (\bibinfo {year} {2011}{\natexlab{a}})}\BibitemShut {NoStop}%
\bibitem [{\citenamefont {{del Campo}}\ \emph {et~al.}(2012)\citenamefont {{del
  Campo}}, \citenamefont {Rams},\ and\ \citenamefont
  {Zurek}}]{delcampo:PRL2012}%
  \BibitemOpen
  \bibfield  {author} {\bibinfo {author} {\bibfnamefont {A.}~\bibnamefont {{del
  Campo}}}, \bibinfo {author} {\bibfnamefont {M.~M.}\ \bibnamefont {Rams}}, \
  and\ \bibinfo {author} {\bibfnamefont {W.~H.}\ \bibnamefont {Zurek}},\
  }\href@noop {} {\bibfield  {journal} {\bibinfo  {journal} {Phys. Rev. Lett.}\
  }\textbf {\bibinfo {volume} {109}},\ \bibinfo {pages} {115703} (\bibinfo
  {year} {2012})}\BibitemShut {NoStop}%
\bibitem [{\citenamefont {Campbell}\ \emph {et~al.}(2014)\citenamefont
  {Campbell}, \citenamefont {{De Chiara}}, \citenamefont {Paternostro},
  \citenamefont {Palma},\ and\ \citenamefont {Fazio}}]{campbell:arXiv2014}%
  \BibitemOpen
  \bibfield  {author} {\bibinfo {author} {\bibfnamefont {S.}~\bibnamefont
  {Campbell}}, \bibinfo {author} {\bibfnamefont {G.}~\bibnamefont {{De
  Chiara}}}, \bibinfo {author} {\bibfnamefont {M.}~\bibnamefont {Paternostro}},
  \bibinfo {author} {\bibfnamefont {G.}~\bibnamefont {Palma}}, \ and\ \bibinfo
  {author} {\bibfnamefont {R.}~\bibnamefont {Fazio}},\ }\href@noop {}
  {\bibfield  {journal} {\bibinfo  {journal} {arXiv:1410.1555}\ } (\bibinfo
  {year} {2014})}\BibitemShut {NoStop}%
\bibitem [{\citenamefont {Bason}\ \emph {et~al.}(2012)\citenamefont {Bason},
  \citenamefont {Viteau}, \citenamefont {N.Malossi}, \citenamefont {Huillery},
  \citenamefont {Arimondo}, \citenamefont {Ciampini}, \citenamefont {Fazio},
  \citenamefont {Giovanetti}, \citenamefont {Mannella},\ and\ \citenamefont
  {Morsch}}]{bason:NatPhys2012}%
  \BibitemOpen
  \bibfield  {author} {\bibinfo {author} {\bibfnamefont {M.~G.}\ \bibnamefont
  {Bason}}, \bibinfo {author} {\bibfnamefont {M.}~\bibnamefont {Viteau}},
  \bibinfo {author} {\bibnamefont {N.Malossi}}, \bibinfo {author}
  {\bibfnamefont {P.}~\bibnamefont {Huillery}}, \bibinfo {author}
  {\bibfnamefont {E.}~\bibnamefont {Arimondo}}, \bibinfo {author}
  {\bibfnamefont {D.}~\bibnamefont {Ciampini}}, \bibinfo {author}
  {\bibfnamefont {R.}~\bibnamefont {Fazio}}, \bibinfo {author} {\bibfnamefont
  {V.}~\bibnamefont {Giovanetti}}, \bibinfo {author} {\bibfnamefont
  {R.}~\bibnamefont {Mannella}}, \ and\ \bibinfo {author} {\bibfnamefont
  {O.}~\bibnamefont {Morsch}},\ }\href@noop {} {\bibfield  {journal} {\bibinfo
  {journal} {Nature Phys.}\ }\textbf {\bibinfo {volume} {8}},\ \bibinfo {pages}
  {147} (\bibinfo {year} {2012})}\BibitemShut {NoStop}%
\bibitem [{\citenamefont {Levy}\ and\ \citenamefont
  {Kosloff}(2012)}]{levy:PRL2012}%
  \BibitemOpen
  \bibfield  {author} {\bibinfo {author} {\bibfnamefont {A.}~\bibnamefont
  {Levy}}\ and\ \bibinfo {author} {\bibfnamefont {R.}~\bibnamefont {Kosloff}},\
  }\href@noop {} {\bibfield  {journal} {\bibinfo  {journal} {Phys. Rev. Lett.}\
  }\textbf {\bibinfo {volume} {108}},\ \bibinfo {pages} {070604} (\bibinfo
  {year} {2012})}\BibitemShut {NoStop}%
\bibitem [{\citenamefont {Schaff}\ \emph {et~al.}(2010)\citenamefont {Schaff},
  \citenamefont {Song}, \citenamefont {Vignolo},\ and\ \citenamefont
  {Labeyrie}}]{schaff:PRA2010}%
  \BibitemOpen
  \bibfield  {author} {\bibinfo {author} {\bibfnamefont {J.~F.}\ \bibnamefont
  {Schaff}}, \bibinfo {author} {\bibfnamefont {X.~L.}\ \bibnamefont {Song}},
  \bibinfo {author} {\bibfnamefont {P.}~\bibnamefont {Vignolo}}, \ and\
  \bibinfo {author} {\bibfnamefont {G.}~\bibnamefont {Labeyrie}},\ }\href@noop
  {} {\bibfield  {journal} {\bibinfo  {journal} {Phys. Rev. A}\ }\textbf
  {\bibinfo {volume} {82}},\ \bibinfo {pages} {033430} (\bibinfo {year}
  {2010})}\BibitemShut {NoStop}%
\bibitem [{\citenamefont {Kagan}\ \emph {et~al.}(1996)\citenamefont {Kagan},
  \citenamefont {Surkov},\ and\ \citenamefont {Shlyapnikov}}]{kagan:PRA1996}%
  \BibitemOpen
  \bibfield  {author} {\bibinfo {author} {\bibfnamefont {Y.}~\bibnamefont
  {Kagan}}, \bibinfo {author} {\bibfnamefont {E.~L.}\ \bibnamefont {Surkov}}, \
  and\ \bibinfo {author} {\bibfnamefont {G.~V.}\ \bibnamefont {Shlyapnikov}},\
  }\href@noop {} {\bibfield  {journal} {\bibinfo  {journal} {Phys. Rev. A}\
  }\textbf {\bibinfo {volume} {54}},\ \bibinfo {pages} {R1753} (\bibinfo {year}
  {1996})}\BibitemShut {NoStop}%
\bibitem [{\citenamefont {Castin}\ and\ \citenamefont
  {Dum}(1996)}]{castin:PRL1996}%
  \BibitemOpen
  \bibfield  {author} {\bibinfo {author} {\bibfnamefont {Y.}~\bibnamefont
  {Castin}}\ and\ \bibinfo {author} {\bibfnamefont {R.}~\bibnamefont {Dum}},\
  }\href@noop {} {\bibfield  {journal} {\bibinfo  {journal} {Phys. Rev. Lett.}\
  }\textbf {\bibinfo {volume} {77}},\ \bibinfo {pages} {5315} (\bibinfo {year}
  {1996})}\BibitemShut {NoStop}%
\bibitem [{\citenamefont {Kagan}\ \emph {et~al.}(1997)\citenamefont {Kagan},
  \citenamefont {Surkov},\ and\ \citenamefont {Shlyapnikov}}]{kagan:PRA1997}%
  \BibitemOpen
  \bibfield  {author} {\bibinfo {author} {\bibfnamefont {Y.}~\bibnamefont
  {Kagan}}, \bibinfo {author} {\bibfnamefont {E.~L.}\ \bibnamefont {Surkov}}, \
  and\ \bibinfo {author} {\bibfnamefont {G.~V.}\ \bibnamefont {Shlyapnikov}},\
  }\href@noop {} {\bibfield  {journal} {\bibinfo  {journal} {Phys. Rev. A}\
  }\textbf {\bibinfo {volume} {55}},\ \bibinfo {pages} {R18} (\bibinfo {year}
  {1997})}\BibitemShut {NoStop}%
\bibitem [{\citenamefont {Muga}\ \emph {et~al.}(2009)\citenamefont {Muga},
  \citenamefont {Chen}, \citenamefont {Ruschhaupt},\ and\ \citenamefont
  {{Guery-Odelin}}}]{muga:JPB2009}%
  \BibitemOpen
  \bibfield  {author} {\bibinfo {author} {\bibfnamefont {J.}~\bibnamefont
  {Muga}}, \bibinfo {author} {\bibfnamefont {X.}~\bibnamefont {Chen}}, \bibinfo
  {author} {\bibfnamefont {A.}~\bibnamefont {Ruschhaupt}}, \ and\ \bibinfo
  {author} {\bibfnamefont {D.}~\bibnamefont {{Guery-Odelin}}},\ }\href@noop {}
  {\bibfield  {journal} {\bibinfo  {journal} {J. Phys. B}\ }\textbf {\bibinfo
  {volume} {42}},\ \bibinfo {pages} {241001} (\bibinfo {year}
  {2009})}\BibitemShut {NoStop}%
\bibitem [{\citenamefont {Schaff}\ \emph {et~al.}(2011)\citenamefont {Schaff},
  \citenamefont {Song}, \citenamefont {Capuzzi}, \citenamefont {Vignolo},\ and\
  \citenamefont {Labeyrie}}]{schaff:EPL2011}%
  \BibitemOpen
  \bibfield  {author} {\bibinfo {author} {\bibfnamefont {J.-F.}\ \bibnamefont
  {Schaff}}, \bibinfo {author} {\bibfnamefont {X.-L.}\ \bibnamefont {Song}},
  \bibinfo {author} {\bibfnamefont {P.}~\bibnamefont {Capuzzi}}, \bibinfo
  {author} {\bibfnamefont {P.}~\bibnamefont {Vignolo}}, \ and\ \bibinfo
  {author} {\bibfnamefont {G.}~\bibnamefont {Labeyrie}},\ }\href@noop {}
  {\bibfield  {journal} {\bibinfo  {journal} {EPL}\ }\textbf {\bibinfo {volume}
  {93}},\ \bibinfo {pages} {23001} (\bibinfo {year} {2011})}\BibitemShut
  {NoStop}%
\bibitem [{\citenamefont {Gritsev}\ \emph {et~al.}(2010)\citenamefont
  {Gritsev}, \citenamefont {Barmettler},\ and\ \citenamefont
  {Demler}}]{gritsev:NJP2010}%
  \BibitemOpen
  \bibfield  {author} {\bibinfo {author} {\bibfnamefont {V.}~\bibnamefont
  {Gritsev}}, \bibinfo {author} {\bibfnamefont {P.}~\bibnamefont {Barmettler}},
  \ and\ \bibinfo {author} {\bibfnamefont {E.}~\bibnamefont {Demler}},\
  }\href@noop {} {\bibfield  {journal} {\bibinfo  {journal} {New J. Phys.}\
  }\textbf {\bibinfo {volume} {12}},\ \bibinfo {pages} {113005} (\bibinfo
  {year} {2010})}\BibitemShut {NoStop}%
\bibitem [{\citenamefont {Rohringer}\ \emph {et~al.}(2013)\citenamefont
  {Rohringer}, \citenamefont {Fischer}, \citenamefont {Steiner}, \citenamefont
  {Mazets}, \citenamefont {Schmiedmayer},\ and\ \citenamefont
  {Trupke}}]{rohringer:arXiv2013}%
  \BibitemOpen
  \bibfield  {author} {\bibinfo {author} {\bibfnamefont {W.}~\bibnamefont
  {Rohringer}}, \bibinfo {author} {\bibfnamefont {D.}~\bibnamefont {Fischer}},
  \bibinfo {author} {\bibfnamefont {F.}~\bibnamefont {Steiner}}, \bibinfo
  {author} {\bibfnamefont {I.}~\bibnamefont {Mazets}}, \bibinfo {author}
  {\bibfnamefont {J.}~\bibnamefont {Schmiedmayer}}, \ and\ \bibinfo {author}
  {\bibfnamefont {M.}~\bibnamefont {Trupke}},\ }\href@noop {} {\bibfield
  {journal} {\bibinfo  {journal} {arXiv:1312.5948}\ } (\bibinfo {year}
  {2013})}\BibitemShut {NoStop}%
\bibitem [{\citenamefont {del Campo}\ and\ \citenamefont
  {Boshier}(2012)}]{delcampo:SciRep2012}%
  \BibitemOpen
  \bibfield  {author} {\bibinfo {author} {\bibfnamefont {A.}~\bibnamefont {del
  Campo}}\ and\ \bibinfo {author} {\bibfnamefont {M.~G.}\ \bibnamefont
  {Boshier}},\ }\href@noop {} {\bibfield  {journal} {\bibinfo  {journal} {Sci.
  Rep.}\ }\textbf {\bibinfo {volume} {2}},\ \bibinfo {pages} {648} (\bibinfo
  {year} {2012})}\BibitemShut {NoStop}%
\bibitem [{\citenamefont {Castin}\ and\ \citenamefont
  {Werner}(2012)}]{castin:Springer2012}%
  \BibitemOpen
  \bibfield  {author} {\bibinfo {author} {\bibfnamefont {Y.}~\bibnamefont
  {Castin}}\ and\ \bibinfo {author} {\bibfnamefont {F.}~\bibnamefont
  {Werner}},\ }in\ \href@noop {} {\emph {\bibinfo {booktitle} {The {BCS}-{BEC}
  Crossover and the Unitary Fermi Gas}}}\ (\bibinfo  {publisher} {Springer},\
  \bibinfo {year} {2012})\BibitemShut {NoStop}%
\bibitem [{\citenamefont {del Campo}(2011{\natexlab{b}})}]{delcampo:EPL2011}%
  \BibitemOpen
  \bibfield  {author} {\bibinfo {author} {\bibfnamefont {A.}~\bibnamefont {del
  Campo}},\ }\href@noop {} {\bibfield  {journal} {\bibinfo  {journal}
  {Europhys. Lett.}\ }\textbf {\bibinfo {volume} {96}},\ \bibinfo {pages}
  {60005} (\bibinfo {year} {2011}{\natexlab{b}})}\BibitemShut {NoStop}%
\bibitem [{\citenamefont {Pitaevskii}(1996)}]{pitaevskii:PhysLettA1996}%
  \BibitemOpen
  \bibfield  {author} {\bibinfo {author} {\bibfnamefont {L.~P.}\ \bibnamefont
  {Pitaevskii}},\ }\href@noop {} {\bibfield  {journal} {\bibinfo  {journal}
  {Phys. Lett. A}\ }\textbf {\bibinfo {volume} {221}},\ \bibinfo {pages} {14}
  (\bibinfo {year} {1996})}\BibitemShut {NoStop}%
\bibitem [{\citenamefont {Pitaevskii}\ and\ \citenamefont
  {Rosch}(1997)}]{pitaevskii:PRA1997}%
  \BibitemOpen
  \bibfield  {author} {\bibinfo {author} {\bibfnamefont {L.~P.}\ \bibnamefont
  {Pitaevskii}}\ and\ \bibinfo {author} {\bibfnamefont {A.}~\bibnamefont
  {Rosch}},\ }\href@noop {} {\bibfield  {journal} {\bibinfo  {journal} {Phys.
  Rev. A}\ }\textbf {\bibinfo {volume} {55}},\ \bibinfo {pages} {R853}
  (\bibinfo {year} {1997})}\BibitemShut {NoStop}%
\bibitem [{\citenamefont {{G}uery{-}{O}delin}\ \emph
  {et~al.}(2014)\citenamefont {{G}uery{-}{O}delin}, \citenamefont {Muga},
  \citenamefont {{R}uiz{-}{M}ontero},\ and\ \citenamefont
  {Trizac}}]{dgo:PRL2014}%
  \BibitemOpen
  \bibfield  {author} {\bibinfo {author} {\bibfnamefont {D.}~\bibnamefont
  {{G}uery{-}{O}delin}}, \bibinfo {author} {\bibfnamefont {J.~G.}\ \bibnamefont
  {Muga}}, \bibinfo {author} {\bibfnamefont {M.~J.}\ \bibnamefont
  {{R}uiz{-}{M}ontero}}, \ and\ \bibinfo {author} {\bibfnamefont
  {E.}~\bibnamefont {Trizac}},\ }\href@noop {} {\bibfield  {journal} {\bibinfo
  {journal} {Phys. Rev. Lett.}\ }\textbf {\bibinfo {volume} {112}},\ \bibinfo
  {pages} {180602} (\bibinfo {year} {2014})}\BibitemShut {NoStop}%
\bibitem [{\citenamefont {Chevy}\ \emph {et~al.}(2002)\citenamefont {Chevy},
  \citenamefont {Bretin}, \citenamefont {Rosenbusch}, \citenamefont {Madison},\
  and\ \citenamefont {Dalibard}}]{chevy:PRL2002}%
  \BibitemOpen
  \bibfield  {author} {\bibinfo {author} {\bibfnamefont {F.}~\bibnamefont
  {Chevy}}, \bibinfo {author} {\bibfnamefont {V.}~\bibnamefont {Bretin}},
  \bibinfo {author} {\bibfnamefont {P.}~\bibnamefont {Rosenbusch}}, \bibinfo
  {author} {\bibfnamefont {K.~W.}\ \bibnamefont {Madison}}, \ and\ \bibinfo
  {author} {\bibfnamefont {J.}~\bibnamefont {Dalibard}},\ }\href@noop {}
  {\bibfield  {journal} {\bibinfo  {journal} {Phys. Rev. Lett.}\ }\textbf
  {\bibinfo {volume} {88}},\ \bibinfo {pages} {250402} (\bibinfo {year}
  {2002})}\BibitemShut {NoStop}%
\bibitem [{\citenamefont {Polkovnikov}\ \emph {et~al.}(2011)\citenamefont
  {Polkovnikov}, \citenamefont {Sengupta}, \citenamefont {Silva},\ and\
  \citenamefont {Vengalattore}}]{polkovnikov_RMP2011}%
  \BibitemOpen
  \bibfield  {author} {\bibinfo {author} {\bibfnamefont {A.}~\bibnamefont
  {Polkovnikov}}, \bibinfo {author} {\bibfnamefont {K.}~\bibnamefont
  {Sengupta}}, \bibinfo {author} {\bibfnamefont {A.}~\bibnamefont {Silva}}, \
  and\ \bibinfo {author} {\bibfnamefont {M.}~\bibnamefont {Vengalattore}},\
  }\href@noop {} {\bibfield  {journal} {\bibinfo  {journal} {Rev. Mod. Phys.}\
  }\textbf {\bibinfo {volume} {83}},\ \bibinfo {pages} {863} (\bibinfo {year}
  {2011})}\BibitemShut {NoStop}%
\bibitem [{\citenamefont {Olshanii}\ \emph {et~al.}(2010)\citenamefont
  {Olshanii}, \citenamefont {Perrin},\ and\ \citenamefont
  {Lorent}}]{olshanii:PRL2010}%
  \BibitemOpen
  \bibfield  {author} {\bibinfo {author} {\bibfnamefont {M.}~\bibnamefont
  {Olshanii}}, \bibinfo {author} {\bibfnamefont {H.}~\bibnamefont {Perrin}}, \
  and\ \bibinfo {author} {\bibfnamefont {V.}~\bibnamefont {Lorent}},\
  }\href@noop {} {\bibfield  {journal} {\bibinfo  {journal} {Phys. Rev. Lett.}\
  }\textbf {\bibinfo {volume} {105}},\ \bibinfo {pages} {095302} (\bibinfo
  {year} {2010})}\BibitemShut {NoStop}%
\bibitem [{\citenamefont {Hu}\ and\ \citenamefont {Liang}(2011)}]{hu:PRL2011}%
  \BibitemOpen
  \bibfield  {author} {\bibinfo {author} {\bibfnamefont {Y.}~\bibnamefont
  {Hu}}\ and\ \bibinfo {author} {\bibfnamefont {Z.}~\bibnamefont {Liang}},\
  }\href@noop {} {\bibfield  {journal} {\bibinfo  {journal} {Phys. Rev. Lett.}\
  }\textbf {\bibinfo {volume} {107}},\ \bibinfo {pages} {110401} (\bibinfo
  {year} {2011})}\BibitemShut {NoStop}%
\bibitem [{\citenamefont {Hofmann}(2012)}]{hofmann:PRL2012}%
  \BibitemOpen
  \bibfield  {author} {\bibinfo {author} {\bibfnamefont {J.}~\bibnamefont
  {Hofmann}},\ }\href@noop {} {\bibfield  {journal} {\bibinfo  {journal} {Phys.
  Rev. Lett.}\ }\textbf {\bibinfo {volume} {108}},\ \bibinfo {pages} {185303}
  (\bibinfo {year} {2012})}\BibitemShut {NoStop}%
\bibitem [{\citenamefont {Merloti}\ \emph {et~al.}(2013)\citenamefont
  {Merloti}, \citenamefont {Dubessy}, \citenamefont {Longchambon},
  \citenamefont {Olshanii},\ and\ \citenamefont {Perrin}}]{merloti:PRA2013}%
  \BibitemOpen
  \bibfield  {author} {\bibinfo {author} {\bibfnamefont {K.}~\bibnamefont
  {Merloti}}, \bibinfo {author} {\bibfnamefont {R.}~\bibnamefont {Dubessy}},
  \bibinfo {author} {\bibfnamefont {L.}~\bibnamefont {Longchambon}}, \bibinfo
  {author} {\bibfnamefont {M.}~\bibnamefont {Olshanii}}, \ and\ \bibinfo
  {author} {\bibfnamefont {H.}~\bibnamefont {Perrin}},\ }\href@noop {}
  {\bibfield  {journal} {\bibinfo  {journal} {Phys. Rev. A}\ }\textbf {\bibinfo
  {volume} {88}},\ \bibinfo {pages} {061603} (\bibinfo {year}
  {2013})}\BibitemShut {NoStop}%
\bibitem [{\citenamefont {Hadzibabic}\ and\ \citenamefont
  {Dalibard}(2011)}]{hadzibabic:NuovoCimento2011}%
  \BibitemOpen
  \bibfield  {author} {\bibinfo {author} {\bibfnamefont {Z.}~\bibnamefont
  {Hadzibabic}}\ and\ \bibinfo {author} {\bibfnamefont {J.}~\bibnamefont
  {Dalibard}},\ }\href@noop {} {\bibfield  {journal} {\bibinfo  {journal} {Riv.
  Nuovo Cimento}\ }\textbf {\bibinfo {volume} {34}},\ \bibinfo {pages} {389}
  (\bibinfo {year} {2011})}\BibitemShut {NoStop}%
\bibitem [{\citenamefont {Petrov}\ and\ \citenamefont
  {Shlyapnikov}(2001)}]{petrov:PRA2001}%
  \BibitemOpen
  \bibfield  {author} {\bibinfo {author} {\bibfnamefont {D.~S.}\ \bibnamefont
  {Petrov}}\ and\ \bibinfo {author} {\bibfnamefont {G.~V.}\ \bibnamefont
  {Shlyapnikov}},\ }\href@noop {} {\bibfield  {journal} {\bibinfo  {journal}
  {Phys. Rev. A}\ }\textbf {\bibinfo {volume} {64}},\ \bibinfo {pages} {012706}
  (\bibinfo {year} {2001})}\BibitemShut {NoStop}%
\bibitem [{\citenamefont {Mora}\ and\ \citenamefont
  {Castin}(2009)}]{mora:PRL2009}%
  \BibitemOpen
  \bibfield  {author} {\bibinfo {author} {\bibfnamefont {C.}~\bibnamefont
  {Mora}}\ and\ \bibinfo {author} {\bibfnamefont {Y.}~\bibnamefont {Castin}},\
  }\href@noop {} {\bibfield  {journal} {\bibinfo  {journal} {Phys. Rev. Lett.}\
  }\textbf {\bibinfo {volume} {102}},\ \bibinfo {pages} {180404} (\bibinfo
  {year} {2009})}\BibitemShut {NoStop}%
\bibitem [{\citenamefont {Cercignani}(1988)}]{cercignani:Springer1988}%
  \BibitemOpen
  \bibfield  {author} {\bibinfo {author} {\bibfnamefont {C.}~\bibnamefont
  {Cercignani}},\ }\href@noop {} {\emph {\bibinfo {title} {The Boltzmann
  Equation and its Applications}}}\ (\bibinfo  {publisher} {Springer},\
  \bibinfo {year} {1988})\BibitemShut {NoStop}%
\bibitem [{Note1()}]{Note1}%
  \BibitemOpen
  \bibinfo {note} {The first three conditions ($b(0)=1$, $\protect \mathaccentV
  {dot}05F{b}(0)=0$, $\protect \mathaccentV {ddot}07F{b}(0)=0$) state the
  stationarity of $\Psi (\protect \bm {r},t)$ for times $t<0$. Conditions 4 and
  5 ($\protect \mathaccentV {dot}05F{b}(t_f)=0$, $\protect \mathaccentV
  {ddot}07F{b}(t_f)=0$) state its stationarity for times $t>t_f$. Condition 6
  follows from Eq.~(\ref {eq:cond_w1fw2f}) and sets the final value of the
  scaling parameter ($b(t_f)=({\omega _{01}/\omega
  _{f1}})^{1/2}$).}\BibitemShut {Stop}%
\bibitem [{\citenamefont {Menotti}\ \emph {et~al.}(2002)\citenamefont
  {Menotti}, \citenamefont {Pedri},\ and\ \citenamefont
  {Stringari}}]{menotti:PRL2002}%
  \BibitemOpen
  \bibfield  {author} {\bibinfo {author} {\bibfnamefont {C.}~\bibnamefont
  {Menotti}}, \bibinfo {author} {\bibfnamefont {P.}~\bibnamefont {Pedri}}, \
  and\ \bibinfo {author} {\bibfnamefont {S.}~\bibnamefont {Stringari}},\
  }\href@noop {} {\bibfield  {journal} {\bibinfo  {journal} {Phys. Rev. Lett.}\
  }\textbf {\bibinfo {volume} {89}},\ \bibinfo {pages} {250402} (\bibinfo
  {year} {2002})}\BibitemShut {NoStop}%
\bibitem [{\citenamefont {{O'Hara}}\ \emph {et~al.}(2002)\citenamefont
  {{O'Hara}}, \citenamefont {Hemmer}, \citenamefont {Gehm}, \citenamefont
  {Granade},\ and\ \citenamefont {Thomas}}]{ohara:Science2002}%
  \BibitemOpen
  \bibfield  {author} {\bibinfo {author} {\bibfnamefont {K.~M.}\ \bibnamefont
  {{O'Hara}}}, \bibinfo {author} {\bibfnamefont {S.~L.}\ \bibnamefont
  {Hemmer}}, \bibinfo {author} {\bibfnamefont {M.~E.}\ \bibnamefont {Gehm}},
  \bibinfo {author} {\bibfnamefont {S.~R.}\ \bibnamefont {Granade}}, \ and\
  \bibinfo {author} {\bibfnamefont {J.~E.}\ \bibnamefont {Thomas}},\
  }\href@noop {} {\bibfield  {journal} {\bibinfo  {journal} {Science}\ }\textbf
  {\bibinfo {volume} {298}},\ \bibinfo {pages} {2179} (\bibinfo {year}
  {2002})}\BibitemShut {NoStop}%
\bibitem [{\citenamefont {Brouzos}\ \emph {et~al.}(2014)\citenamefont
  {Brouzos}, \citenamefont {Streltsov}, \citenamefont {Negretti}, \citenamefont
  {Said}, \citenamefont {Canneva}, \citenamefont {Montangero},\ and\
  \citenamefont {Calarco}}]{brouzos:arXiv2014}%
  \BibitemOpen
  \bibfield  {author} {\bibinfo {author} {\bibfnamefont {I.}~\bibnamefont
  {Brouzos}}, \bibinfo {author} {\bibfnamefont {A.~I.}\ \bibnamefont
  {Streltsov}}, \bibinfo {author} {\bibfnamefont {A.}~\bibnamefont {Negretti}},
  \bibinfo {author} {\bibfnamefont {R.~S.}\ \bibnamefont {Said}}, \bibinfo
  {author} {\bibfnamefont {T.}~\bibnamefont {Canneva}}, \bibinfo {author}
  {\bibfnamefont {S.}~\bibnamefont {Montangero}}, \ and\ \bibinfo {author}
  {\bibfnamefont {T.}~\bibnamefont {Calarco}},\ }\href@noop {} {\bibfield
  {journal} {\bibinfo  {journal} {arXiv:1412.6142}\ } (\bibinfo {year}
  {2014})}\BibitemShut {NoStop}%
\bibitem [{\citenamefont {He}\ \emph {et~al.}(2012)\citenamefont {He},
  \citenamefont {Yu}, \citenamefont {Xu}, \citenamefont {Wang},\ and\
  \citenamefont {Zhan}}]{he:OptExpress2012}%
  \BibitemOpen
  \bibfield  {author} {\bibinfo {author} {\bibfnamefont {X.}~\bibnamefont
  {He}}, \bibinfo {author} {\bibfnamefont {S.}~\bibnamefont {Yu}}, \bibinfo
  {author} {\bibfnamefont {P.}~\bibnamefont {Xu}}, \bibinfo {author}
  {\bibfnamefont {J.}~\bibnamefont {Wang}}, \ and\ \bibinfo {author}
  {\bibfnamefont {M.}~\bibnamefont {Zhan}},\ }\href@noop {} {\bibfield
  {journal} {\bibinfo  {journal} {Opt. Express}\ }\textbf {\bibinfo {volume}
  {20}} (\bibinfo {year} {2012})}\BibitemShut {NoStop}%
\end{thebibliography}

\end{document}